# Observation of a phase transition within the domain walls of ferromagnetic $Co_3Sn_2S_2$

Changmin Lee[1], Praveen Vir[2], Kaustuv Manna[2,3], Chandra Shekhar[2], J. E. Moore[1,4], M. A. Kastner[5,6,7], Claudia Felser[2], and Joseph Orenstein[1,4*]

[1] Materials Science Division, Lawrence Berkeley National Laboratory, Berkeley, California, USA

[2] Max Planck Institute for Chemical Physics of Solids, Dresden, Germany

[3] Department of Physics, Indian Institute of Technology Delhi, New Delhi, India

[4] Department of Physics, University of California at Berkeley, Berkeley, California, USA

[5] Department of Physics, Stanford University, Palo Alto, California, USA

[6] Stanford Institute for Materials and Energy Science, SLAC National Accelerator Laboratory, Menlo Park, California, USA

[7] Department of Physics, Massachusetts Institute of Technology, Cambridge, MA, USA

* Corresponding author: jworenstein@lbl.gov

## Abstract

**The ferromagnetic phase of $Co_3Sn_2S_2$ is widely considered to be a topological Weyl semimetal, with evidence for momentum-space monopoles of Berry curvature from transport and spectroscopic probes. As the bandstructure is highly sensitive to the magnetic order, attention has focused on anomalies in magnetization, susceptibility and transport measurements that are seen well below the Curie temperature, leading to speculation that a "hidden" phase coexists with ferromagnetism. Here we report spatially-resolved measurements by Kerr effect microscopy that identify this phase. We find that the anomalies coincide with a deep minimum in domain wall (DW) mobility, indicating a crossover between two regimes of DW propagation. We demonstrate that this crossover is a manifestation of a 2D phase transition that occurs within the DW, in which the magnetization texture changes from continuous rotation to unidirectional variation. We propose that the existence of this 2D transition deep within the ferromagnetic state of the bulk is a consequence of a giant quality factor for magnetocrystalline anisotropy unique to this compound. This work broadens the horizon of the conventional binary classification of DWs into Bloch and Néel** **walls, and suggests new strategies for manipulation of domain walls and their role in electron and spin transport.**

## Introduction

The crystal structure of $Co_3Sn_2S_2$ comprises quasi-2D $Co_3Sn$ layers separated by sulfur atoms, with the magnetic Co atoms arranged on a kagomé lattice[1]. The ferromagnetic state is characterized by strong easy-axis anisotropy favoring magnetization in the $\boldsymbol{c}$ direction, that is, perpendicular to the layers. Recent research has focused on evidence for Weyl semimetal topology[2-4] in $Co_3Sn_2S_2$ based on a giant anomalous Hall effect[5-9], and observation of Fermi arcs by photoemission[10] and scanning tunneling spectroscopies[11]. However, as the bandstructure is expected to be highly sensitive to the magnetization texture, attention has been drawn to unexpected structures in the temperature dependence near 130 K (well below the Curie temperature of 175 K) that is seen by static and ac susceptibility[12], muon spin rotation[13], Hall effect[14], magneto-optic spectroscopy[8], and magnetic force microscopy[15] measurements. Although these structures are relatively small in some cases, their ubiquity across laboratories and probes has led to proposals that a phase transition to a coexisting antiferromagnet[13] or spin glass[14] takes place deep within the dominant ferromagnetic phase. Here the scanning Kerr microscopy measurements described below show that such a phase transition does indeed occur, but within the 2D manifold of the magnetic domain walls (DWs) rather than the bulk of the 3D crystal.

## Results

A schematic of the experimental setup is shown in Fig. 1(a), which illustrates local probing of magnetization dynamics using the polar Kerr effect (see Methods for details). Scanning the sample under the laser focus yields a map (Fig. 1(b)) of the Kerr ellipticity, $\Phi(\boldsymbol{r})$, revealing domains of magnetization directed parallel and anti-parallel to the surface normal, or $c$-axis direction. The change in $\Phi(\boldsymbol{r})$ across a DW is proportional to the zero-field magnetization (shown as a function of temperature in Fig. S1 of Supplementary Information). For local measurements of the DW dynamics, a coil surrounding the sample generates an oscillating magnetic field parallel to the easy axis, $\boldsymbol{H}(t) = \hat{\boldsymbol{c}} H_{ac} sin\omega t$. This ac magnetic field tilts the energy landscape, inducing DW displacement in the direction that increases the volume of magnetization aligned parallel to $\boldsymbol{H}(t)$. The oscillating displacement in turn generates a synchronous modulation in ellipticity, $\delta\Phi_{\mathrm{ac}}(\boldsymbol{r})$, that peaks at the domain boundaries. (Fig. S2(b) of Supplementary Information is a map of $\delta\Phi_{\mathrm{ac}}(\boldsymbol{r})$ that shows peak contrast at the domain walls imaged in Fig. 1(b)).

Figure 1(c)-(j) shows a series of maps that illustrate the temperature dependence of $\delta\Phi_{\mathrm{ac}}(\boldsymbol{r})/\Phi_{dc}$ in the range from 120 K to the Curie temperature. Two features are immediately clear from the sequence of images. First, as the temperature is increased to the Curie temperature, each domain boundary becomes increasingly convoluted, suggesting a progressive decrease in surface tension. A second, and less expected feature is that the DWs are nearly invisible in the 140 K map, indicating a deep minimum in $\delta\Phi_{\mathrm{ac}}(\boldsymbol{r})/\Phi_{dc}$, and therefore the oscillating displacement $\Delta x$, as a function of $T$.

The key advantage of scanned local probe over global ac susceptibility ($\chi_{ac}$) measurements is the ability to determine the displacement, $\Delta x$, of individual segments of DWs. The displacement is obtained with ≈ 1 nm sensitivity from the ratio of $\delta\Phi_{ac}$ at the domain wall center, $\delta\Phi_{ac}(0)$, to the step in ellipticity across the DW, $\Phi_{dc}$, through the relation $\delta\Phi_{ac}(0)/\Phi_{dc} = \mathrm{erf}(\Delta x/\sqrt{2}\sigma)$, where $\sigma$ is the probe focal radius (see Supplementary Information Section III). Panels (a)-(d) of Figure 2 illustrate the local dynamics of a representative segment of a DW (see Supplementary Information Section IV for details on the spatial variation of DW dynamics). Fig. 2(a) presents an overview of $\delta\Phi_{ac}(0)/\Phi_{dc}$ in the $\omega - T$ plane using a color scale; Fig. 2(b) shows several line cuts through the plane at constant $\omega$. A deep minimum in $\delta\Phi_{ac}(0)/\Phi_{dc}$ is evident in both plots. Fig. 2(c) shows the corresponding DW displacement as a function of $H_{ac}$ for several temperatures. The displacement displays a threshold, indicating collective pinning behavior; $\Delta x(H_{ac})$ is nearly zero below a frequency dependent field, $H_{th}$, and increases linearly with $H - H_{th}$ for $H > H_{th}$. Finally, the wall displacement measured with $H_{ac}$ = 28 Oe is plotted in Fig. 2d on an expanded temperature scale that highlights the step-like change centered on $T$ =135 K.

The step change in displacement near 135 K cannot be understood within the standard theory of DW dynamics, which assumes a Bloch wall texture in which $\boldsymbol{M}$ rotates with constant magnitude in the plane of the DW. In this theory Bloch DW mobility is proportional to the product of the Gilbert parameter, which describes the damping rate of transverse magnetization fluctuations, and the DW width[16]. Both of these parameters are expected to vary smoothly and monotonically at temperatures well below $T_c$. While the disorder that gives rise to collective pinning could account for a monotonic decrease in mobility as $T$ is lowered, it cannot explain the subsequent increase of DW displacement below $\approx 135$ K.

Studies of the *ac* susceptibility, $\chi_\omega(T)$, of insulating hexaferrite magnets performed in the early 1990's provide a clue as to the origin of the anomalous DW displacement in $Co_3Sn_2S_2$. These measurements revealed nonmonotonicity in $\chi_\omega(T)$, resembling the minimum in DW displacement in $Co_3Sn_2S_2$, but with the qualitative difference that it was found in the critical regime, $T = 0.995T_c$, rather that deep within the ferromagnetic phase[17-19]. This phenomenon was traced to the prediction by Bulaevskii and Ginzburg (BG)[20] that a "linear" wall, in which $\boldsymbol{M}$ is aligned along the easy axis and passes through zero at the DW center, has lower energy than the Bloch wall, above a transition temperature, $T_{BG}$. Later the same underlying physics was rediscovered in the context of a transition from sine-Gordon to $\phi^4$ solitons[21,22] as a function of the amplitude of rotational symmetry breaking.

The origin of the BG phase transition and the associated minimum in $\chi_\omega(T)$ is a crossover between the energy required to twist the magnetization and the energy required to change its amplitude. This competition can be quantified by a free energy of the form,

$$F \propto \rho_s(\nabla M)^2 + KM_T^2 + \frac{1}{8\chi_c}(M^2 - M_s^2)^2/M_s^2, \tag{1}$$

where $\rho_s$ is the spin stiffness, $K \equiv 1/2\chi_{ab}$ is the uniaxial (easy axis) anisotropy factor, and $\chi_c$ is a susceptibility that parameterizes the energy cost of changing the amplitude of the magnetization from its equilibrium value. $M_T$ and $M_s$ are the in-plane (or transverse) and saturation magnetizations, respectively. As $T \to T_c$, $\chi_c(T)$ diverges, whereas $K(T)$ does not, and therefore variations in the amplitude of $\boldsymbol{M}$ cost less energy than rotations. Consequently, in a range of temperature below $T_c$ that depends on parameters, the energy to create a linear wall is less than the energy of a Bloch wall. Minimizing the free energy in Eq. 1 yields a continuous phase transition from linear to elliptical to circular (Bloch) walls. The order parameter of the BG transition is $M_T(0)$, the amplitude of the magnetization transverse to the propagation direction evaluated at the center of the DW. This order parameter manifests in DW dynamics because propagation of a linear wall in which $M_T(0) = 0$ involves purely longitudinal changes in magnetization, whereas motion of a Bloch wall, where $M_T(0) = M_s$, requires rotation of $\boldsymbol{M}$.

The signature of such a continuous phase transition is enhanced susceptibility of the order parameter to its conjugate field, in this case a static magnetic field $H_T$ applied perpendicular to the easy axis. Figure 3(a) shows the DW displacement vs. temperature for $H_T = 0, 500$ Oe and 1000 Oe; Fig. 3(b) is the corresponding change in DW displacement with respect to $H_T = 0$. Clearly, the susceptibility to in-plane field is confined to a temperature regime close to the transition. Figs. 3(c) and 3(d) provide a more detailed view of the field dependence of the wall displacement. Fig. 3(c) shows $\Delta x$ $vs.$ $H_{ac}$ for several values of $H_T$ at $T = 139$ $K$; Fig. 3(d) is a plot of $\Delta x$ $vs.$ $H_T$ at $H_{ac} = 28$ Oe for temperatures near the transition. Near the transition temperature a transverse field of 1000 Oe is sufficient to switch the DW from its low to high mobility state, that is, changing $M_T(0)$ from zero to $M_s$. This indicates a remarkable enhancement of transverse susceptibility, as the field required to overcome the bulk anisotropy and rotate the spins to the basal plane is ~ 20 T.

The free energy in Eq. 1 predicts a mean-field transition occurring when the control parameter $\tau(T) \equiv 4\chi_c/\chi_{ab}$ reaches unity and an order parameter critical exponent of one-half, such that. $\rho \equiv M_T(0)/M_s = [1 - \tau(T)]^{1/2}$. However, given that the transition occurs in a two-dimensional (2D) manifold, one might anticipate strong fluctuation effects. Indeed, Lawrie and Lowe have shown theoretically that the BG transition is in the universality class of 2D Ising models in which fluctuations reduce the transition temperature and lead to an order parameter critical exponent of 1/8[23]. The Ising nature of the transition reflects the two discrete choices for the helicity of the Bloch wall.

To test the prediction of a 2D Ising transition within the DW, we determined $\tau(T)$ through measurements of $M$ $vs.H$ for both in- and out-of-plane directions at intervals of 1 K in the temperature range from 110 to 160 K. An example of the isothermal magnetization data at $T =$ 120 K, characteristic of a ferromagnet with strong uniaxial anisotropy, is shown in Fig. 4(a). The in-plane susceptibility $\chi_{ab}$ is extracted from the slope of $M_a$ vs. $H_a$and is related to the anisotropy parameter through the relation $K = 1/2\chi_{ab}$. The longitudinal susceptibility $\chi_c$ is obtained from the slope of $M_c$ vs. $H_c$ just above the saturation point (between 0.2 and 1 T). The temperature dependence of $\chi_c$ and $\chi_{ab}$ (divided by 8 for clarity) is plotted in Fig. 4b.

Fig. 4(c) shows the local DW dynamics over the same expanded temperature range as in Fig. 2(d), but now plotted as a function of $\tau(T)$ rather than $T$. To compare with a theory for the dynamics to be discussed below, we have converted the vertical axis scale from displacement, $\Delta x$, to an effective mobility, $\mu_{\text{eff}}$, through the relation, $\mu_{\text{eff}} \equiv \omega\Delta x/(H_{ac} - H_{th})$. The observation that the transition takes place when $\tau(T)$ is of order unity gives additional strong support for the identifying the step in mobility as the BG transition. Moreover, the fact that the transition occurs at $\tau(T) \approx 0.45$ rather than precisely unity suggests a reduction critical temperature typical of 2D Ising models, where for example the $T_c$ for the Ising transition on a square lattice is reduced from the mean field value by a factor 0.57[24].

To further pursue the mean field vs. 2D Ising comparison, we consider the rate at which the order parameter grows below the transition, using the mobility as a proxy. To do so requires a phenomenological model for the mobility as a function the BG order parameter[22]. We introduce the coefficients $\Gamma_L$ and $\Gamma_T$, the relaxation rate of longitudinal and transverse fluctuations of $\boldsymbol{M}$, respectively, which quantify the rate at which energy is dissipated as the DW propagates. The DW mobility is obtained from the steady state condition that the dissipation rate equals the rate at which energy is gained as the magnetization aligns with the applied field (see Supplementary Section VI for details). The resulting expression for $\mu_{\text{eff}}[\rho(\tau)]$ interpolates between the mobilities of perfectly circular walls where $\mu_{\text{eff}} = w\Gamma_T/M_s$ and linear walls for which $\mu_{\text{eff}} = 3w\Gamma_L/2M_s$. The dashed blue line in Fig. 4(c), which is a plot of $\mu_{\text{eff}}(\tau)$ assuming $\rho = (1-\tau)^{1/2}$ , shows that a mean-field description fails not only for the critical value of $\tau$, but the order parameter exponent as well. The mean field or 3D Ising fits with lower critical values of $\tau$ near 0.5 also fail to describe the data accurately near the phase transition (see Supplementary Section VII for details).
The red dashed line, which assumes the 2D Ising critical exponent of 1/8 and a reduced critical value of $\tau$, such that $\rho = (1-\tau/0.45)^{1/8}$ , provides a better description of the mobility data.

## Discussion

Finally, a natural question is why a DW phase transition deep in the FM state is uniquely observed in $Co_3Sn_2S_2$ and not in other metallic uniaxial ferromagnets. We suggest that what sets this compound apart is the giant value of its dimensionless anisotropy factor, $K$. Table 1 summarizes the magnetocrystalline anisotropy of several uniaxial ferromagnets that are receiving attention for potential applications. Notice that while the dimensionful anisotropy energy, $\mu_0 K M_s^2$, is similar for each compound, the saturation magnetization of $Co_3Sn_2S_2$ is much smaller; consequently, the ratio of anisotropy to magnetostatic energy is anomalously large. This leads to the absence of the domain branching that complicates the structure of domain walls at the surface and disrupts the BG transition in other FM systems. The large value of $K$ likely originates from the extreme sensitivity of the low-energy electronic states to the magnetization[25]. The giant anisotropy in $Co_3Sn_2S_2$, and potentially other systems in which the electron dispersion is highly sensitive to magnetization texture, may lead to new strategies for manipulation of domain walls and their role in electron and spin transport.

## Methods

### Scanning MOKE microscopy

A linearly polarized 633 nm HeNe laser beam was focused at normal incidence onto a 1 μm spot on the sample surface with an objective lens (Olympus LMPFLN 50x, NA = 0.5). The reflected beam was then collected by a 50:50 non-polarizing beam splitter, which directs the beam to a quarter wave plate and a Wollaston prism for balanced photodetection of Kerr ellipticity. The sample position was raster scanned with *xy* piezoelectric scanners (Attocube ANPx101) and the focusing was fine-adjusted with a *z* piezoelectric scanner (Attocube ANPz102). For local measurements of the DW dynamics, an out-of-plane AC magnetic field was applied through a coil (Woodruff Scientific, 156 turns, inner diameter: 5mm, height: 1.5 mm) that surrounds the sample. The modulation in Kerr ellipticity $\delta\Phi_{\mathrm{ac}}(\boldsymbol{r})$ was measured using a standard lock-in detection at the frequency of the coil. The sensitivity of DC Kerr measurements is typically ~1 mrad, due to various sources of $1/f$ fluctuations. By shifting the observation frequency using modulation of $H_{ac}$, we are able to measure Kerr effect amplitudes of less than 1 μrad.

**Data availability Statement**

The datasets generated during and/or analysed during the current study are available from the corresponding author on reasonable request.

## References

1 Weihrich, R. & Anusca, I. Half Antiperovskites. III. Crystallographic and Electronic Structure Effects in Sn2−xInxCo3S2. *Zeitschrift für anorganische und allgemeine Chemie* **632**, 1531-1537, doi:10.1002/zaac.200500524 (2006).

2 Wan, X., Turner, A. M., Vishwanath, A. & Savrasov, S. Y. Topological semimetal and Fermi-arc surface states in the electronic structure of pyrochlore iridates. *Physical Review B* **83**, doi:10.1103/PhysRevB.83.205101 (2011).

3 Burkov, A. A. & Balents, L. Weyl semimetal in a topological insulator multilayer. *Phys Rev Lett* **107**, 127205, doi:10.1103/PhysRevLett.107.127205 (2011).

4 Armitage, N. P., Mele, E. J. & Vishwanath, A. Weyl and Dirac semimetals in three-dimensional solids. *Reviews of Modern Physics* **90**, doi:10.1103/RevModPhys.90.015001 (2018).

5 Xu, Q. *et al.* Topological surface Fermi arcs in the magnetic Weyl semimetal Co3Sn2S2. *Physical Review B* **97**, doi:10.1103/PhysRevB.97.235416 (2018).

6 Liu, E. *et al.* Giant anomalous Hall effect in a ferromagnetic Kagome-lattice semimetal. *Nat Phys* **14**, 1125-1131, doi:10.1038/s41567-018-0234-5 (2018).

7 Wang, Q. *et al.* Large intrinsic anomalous Hall effect in half-metallic ferromagnet Co3Sn2S2 with magnetic Weyl fermions. *Nat Commun* **9**, 3681, doi:10.1038/s41467-018-06088-2 (2018).

8 Okamura, Y. *et al.* Giant magneto-optical responses in magnetic Weyl semimetal Co3Sn2S2. *Nat Commun* **11**, 4619, doi:10.1038/s41467-020-18470-0 (2020).

9 Yang, H. *et al.* Giant anomalous Nernst effect in the magnetic Weyl semimetal $Co_3Sn_2S_2$. *Physical Review Materials* **4**, doi:10.1103/PhysRevMaterials.4.024202 (2020).

10 Liu, D. F. *et al.* Magnetic Weyl semimetal phase in a Kagome crystal. *Science* **365**, 1282-1285, doi:10.1126/science.aav2873 (2019).

11 Morali, N. *et al.* Fermi-arc diversity on surface terminations of the magnetic Weyl semimetal Co3Sn2S2. *Science* **365**, 1286-1291, doi:10.1126/science.aav2334 (2019).

12 Kassem, M. A., Tabata, Y., Waki, T. & Nakamura, H. Low-field anomalous magnetic phase in the kagome-lattice shandite Co3Sn2S2. *Physical Review B* **96**, doi:10.1103/PhysRevB.96.014429 (2017).

13 Guguchia, Z. *et al.* Tunable anomalous Hall conductivity through volume-wise magnetic competition in a topological kagome magnet. *Nat Commun* **11**, 559, doi:10.1038/s41467-020-14325-w (2020).

14 Lachman, E. *et al.* Exchange biased anomalous Hall effect driven by frustration in a magnetic kagome lattice. *Nat Commun* **11**, 560, doi:10.1038/s41467-020-14326-9 (2020).

15 Howlader, S., Ramachandran, R., Monga, S., Singh, Y. & Sheet, G. Domain structure evolution in the ferromagnetic Kagome-lattice Weyl semimetal Co3Sn2S2. *J Phys Condens Matter*, doi:10.1088/1361-648X/abc4d1 (2020).

16 Schryer, N. L. & Walker, L. R. The motion of 180° domain walls in uniform dc magnetic fields. *Journal of Applied Physics* **45**, 5406-5421, doi:10.1063/1.1663252 (1974).

17 Kotzler, J., Garanin, D. A., Hartl, M. & Jahn, L. Evidence for critical fluctuations in Bloch walls near their disordering temperature. *Phys Rev Lett* **71**, 177-180, doi:10.1103/PhysRevLett.71.177 (1993).

18 Kötzler, J., Hartl, M. & Jahn, L. Signature of linear (Bulaevskii–Ginzburg) domain walls near Tc of Ba-hexaferrite. *Journal of Applied Physics* **73**, 6263-6265, doi:10.1063/1.352664 (1993).

19 Hartl-Malang, M., Kotzler, J. & Garanin, D. A. Domain-wall relaxation near the disorder transition of Bloch walls in Sr hexaferrite. *Phys Rev B Condens Matter* **51**, 8974-8983, doi:10.1103/physrevb.51.8974 (1995).

20 Bulaevskii, L. N. & Ginzburg, V. L. Temperature Dependence of the Shape of the Domain Wall in Ferromagnetics and Ferroelectrics. *Sov. Phys. JETP* **18** (1964).

21 Sarker, S., Trullinger, S. E. & Bishop, A. R. Solitary-wave solution for a complex one-dimensional field. *Physics Letters A* **59**, 255-258, doi:10.1016/0375-9601(76)90784-2 (1976).

22 Garanin, D. A. Dynamics of elliptic domain walls. *Physica A: Statistical Mechanics and its Applications* **178**, 467-492, doi:10.1016/0378-4371(91)90033-9 (1991).

23 Lawrie, I. D. & Lowe, M. J. Domain wall singularity in a Landau-Ginzburg model. *Journal of Physics A: Mathematical and General* **14**, 981-992, doi:10.1088/0305-4470/14/4/025 (1981).

24 Onsager, L. Crystal Statistics. I. A Two-Dimensional Model with an Order-Disorder Transition. *Physical Review* **65**, 117-149, doi:10.1103/PhysRev.65.117 (1944).

25 Ghimire, M. P. *et al.* Creating Weyl nodes and controlling their energy by magnetization rotation. *Physical Review Research* **1**, doi:10.1103/PhysRevResearch.1.032044 (2019).

26 Chen, B. *et al.* Magnetic Properties of Layered Itinerant Electron Ferromagnet Fe3GeTe2. *Journal of the Physical Society of Japan* **82**, 124711, doi:10.7566/jpsj.82.124711 (2013).

27 Dillon, J. F. & Olson, C. E. Magnetization, Resonance, and Optical Properties of the Ferromagnet CrI3. *Journal of Applied Physics* **36**, 1259-1260, doi:10.1063/1.1714194 (1965).

## Acknowledgements

We would like to thank Jeffrey Orenstein and Simon Woodruff for design of the magnetic coils used in the measurements, Veronika Sunko and Dylan Rees for assistance in installing the DC magnetic field module, and Jonathan Denlinger for assistance in sample characterization. This work was primarily supported by the Spin Physics program under the Director, Office of Science, Office of Basic Energy Sciences, Materials Sciences and Engineering Division, of the U.S. Department of Energy, Contract No. DE-AC02-76SF00515. C. L. and J.O. acknowledge support from the Quantum Materials program under the Director, Office of Science, Office of Basic Energy Sciences, Materials Sciences and Engineering Division, of the U.S. Department of Energy, Contract No. DE-AC02-05CH11231. J.O. received support for optical measurements from the Gordon and Betty Moore Foundation's EPiQS Initiative through Grant GBMF4537 to J.O. at UC Berkeley. P.V., K.M., C.S., and C.F. acknowledge financial support from the European Research Council (ERC) Advanced Grant No. 742068 "TOP-MAT"; European Union's Horizon 2020 research and innovation program (Grant Nos. 824123 and 766566) and Deutsche Forschungsgemeinschaft (DFG) through SFB 1143. K. M. acknowledges the Max Planck Society for the funding support under Max Planck–India partner group project.

## Author Contributions

C.L. performed the optical measurements. C.L. and J.O. analyzed the data and wrote the manuscript. P.V., K.M., C.S., and C.F. grew and characterized the crystals. C.S. performed the magnetic susceptibility measurements. J.O. and J.E.M. developed the theoretical models. All authors commented on the manuscript.

## Competing interests Statement

The authors declare no competing interests.

| Compound | Magnetization ($\boldsymbol{M_s}$) $\times 10^5 A/m$ | Magnetostatic energy ($\mu_0 M_s^2$) $\times 10^5 J/m^3$ | Anisotropy energy ($\mu_0 K M_s^2$) $\times 10^5 J/m^3$ | Anisotropy factor ($\boldsymbol{K}$) |
|---|---|---|---|---|
| $Co_3Sn_2S_2$[10] | 0.56 | 0.042 | 3.8 | 90 |
| $Fe_3Ge_2Te_2$[26] | 4.0 | 2.0 | 12.1 | 6.0 |
| $CrI_3$[27] | 2.1 | 0.58 | 3.1 | 5.3 |

**Table 1 | Summary of various magnetic constants in uniaxial ferromagnets**

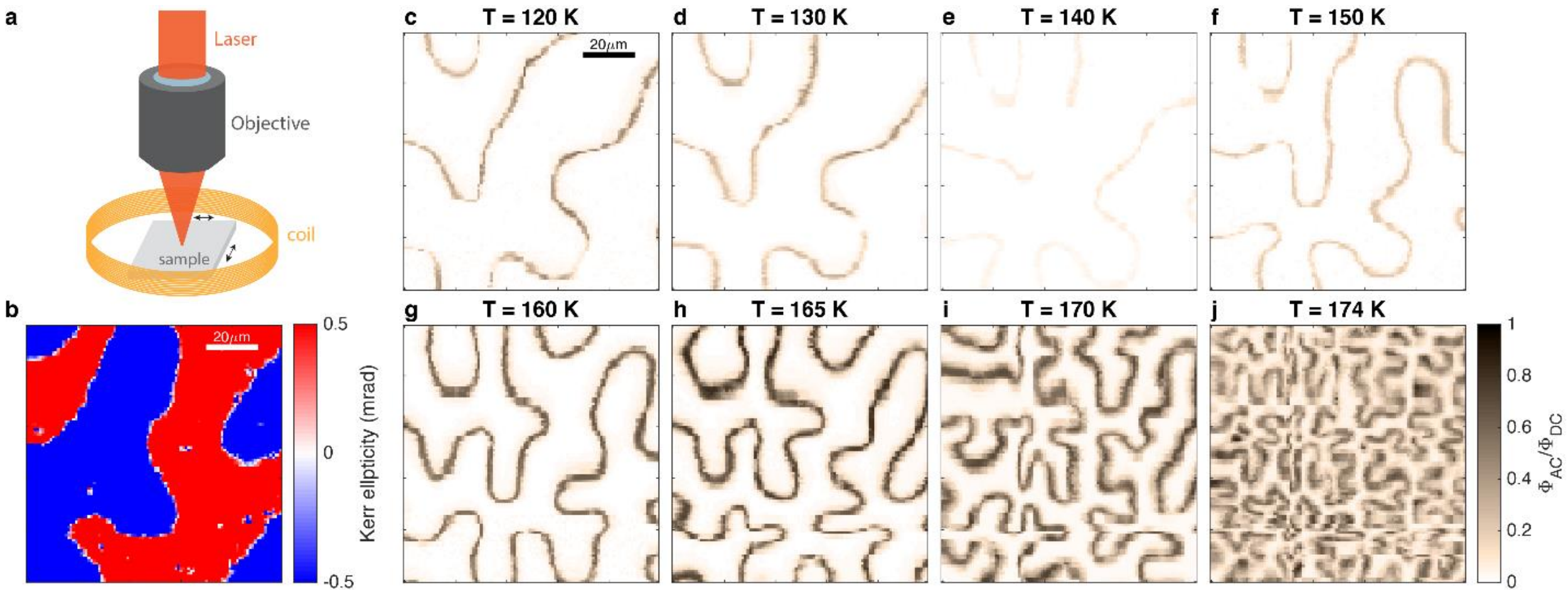

**Figure 1 | Scanning Kerr microscopy. a** Overview of ac MOKE microscopy setup. The sample is surrounded by a coil that generates an out-of-plane ac magnetic field. **b** Unmodulated Kerr ellipticity map taken at T = 120 K reveals stripe-like magnetic domains. **c-j** ac MOKE maps measured at temperatures ranging from 120 to 174 K. A 28 Oe ac magnetic field was applied at a frequency of 1kHz. The normalized ac Kerr ellipticity amplitude $\Phi_{AC}/\Phi_{DC}$ is significantly reduced at 140 K.

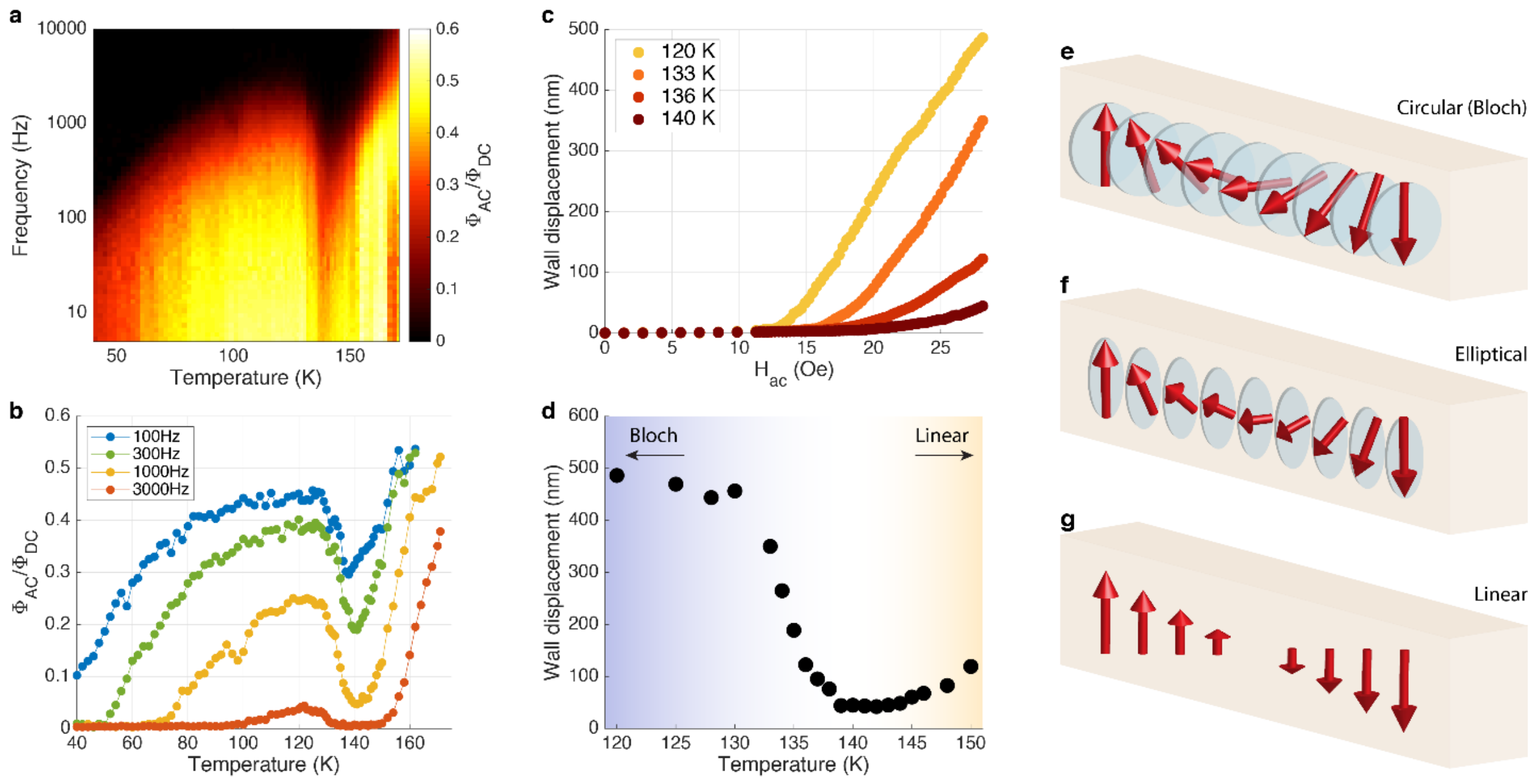

**Figure 2 | DW displacement and linear wall phase transition. a** Normalized ac Kerr response measured at a segment of a domain wall over a range of frequency (5 Hz – 10 kHz) and temperature (40 – 175 K). A sharp decrease in the Kerr response is noticeable near T=140 K. **b** Selected plots of the temperature dependence at constant frequency (horizontal cuts through the frequency-temperature map shown in a). **c** Measurements of the domain wall displacement plotted against ac magnetic field amplitude at temperatures in the range of 120-140 K ( $f = 1$ kHz , See Supplementary Fig. 4 for data at higher temperatures). **d** DW displacement versus temperature. The non-monotonic temperature dependence of the displacement can be understood as a phase transition from a **e**, circular or **f**, elliptical wall at low temperatures to a **g**, linear wall at high temperatures in which the magnetization vanishes at the wall center.

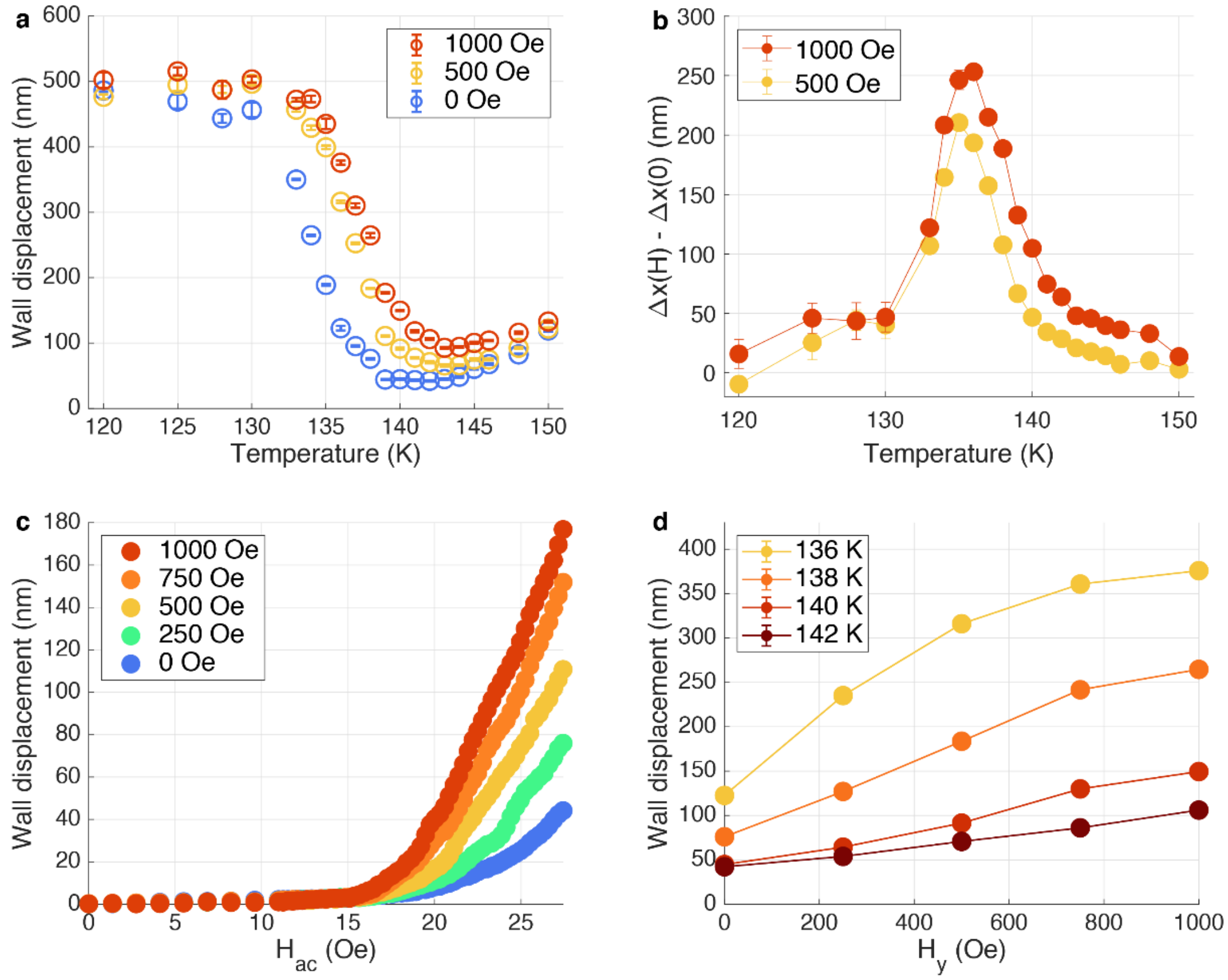

**Figure 3 | Magnetic field dependence of the DW ellipticity. a**. DW displacement plotted against temperature for various in-plane dc magnetic field $H_y$ applied to the sample. **b**. The changes in DW displacement $\Delta x$ induced by dc field values of $H_y = 500$ Oe and $1000$ Oe show sharp peaks at $T = 135\ K$ and $136\ K$, indicating a phase transition into the linear wall state, **c**. DW displacement versus $H_{\mathrm{ac}}$ for different values of in-plane dc fields $H_y$ at $T = 139$ K. A clear change in the DW displacement and mobility is observed by tuning the wall ellipticity with an in-planed field $H_y$ applied to the sample near $T_{\mathrm{BG}}$. **d**. DW displacement plotted versus in-plane field $H_y$ for various temperatures. A saturation behavior is observed at lower temperatures toward the Bloch wall phase.

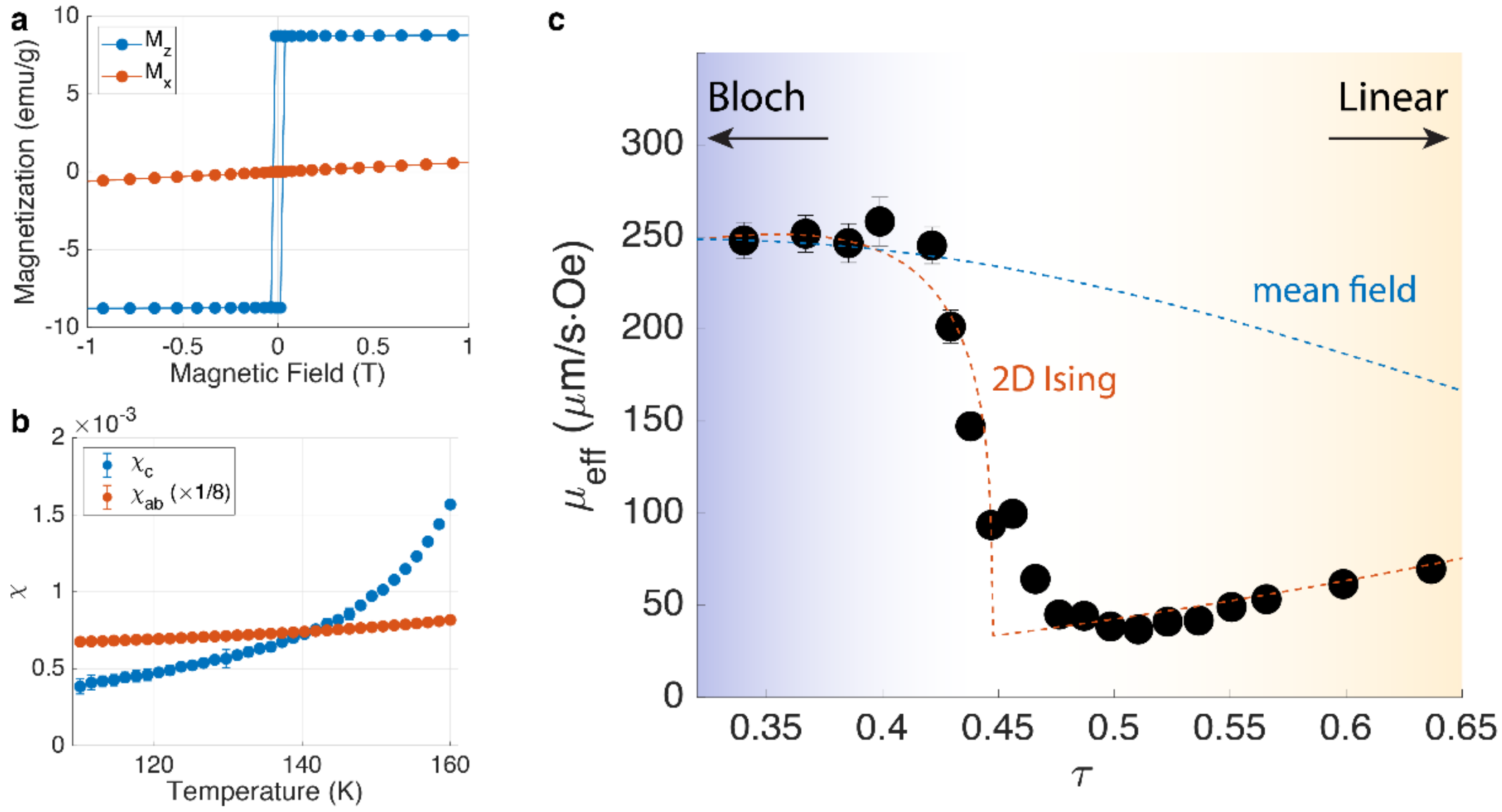

**Figure 4 | Magnetic susceptibility and 2D Ising transition. a** Magnetization versus magnetic field along out-of-plane ($z$) and in-plane ($x$) directions, respectively. **b** Magnetic susceptibility plotted against temperature. The in-plane susceptibility data divided by 8 for clarity. **c** Effective mobility, $\mu_{\mathrm{eff}} = \frac{\Delta x}{\Delta H}$, of the DW is plotted against $\tau \equiv 4\chi_c/\chi_{ab}$, which is the control parameter for the linear wall phase transition. The $\mu_{\mathrm{eff}}$ data fits is consistent with the DW mobility predicted by the 2D Ising model (red dotted line)[24], but not with the mean field theory (blue dotted line), which predict a phase transition at $\tau^{*}\sim 1$. Error bars represent the uncertainties of least square fitting of data.

## Supplementary Information

## "Observation of a phase transition within the domain walls of ferromagnetic $Co_3Sn_2S_2$"

### I. DC Kerr ellipticity on traversing a domain wall

The unmodulated, dc Kerr ellipticity was measured as a function of position transverse to the domain walls. Typically, three consecutive scans through a wall were recorded to improve signal to noise ratio (Fig. S1(a)). The results were then fitted to an error function whose amplitude is proportional to the $z$ component of the magnetization, $M_z$, within a domain. $\Phi_{dc}$ is therefore a measure of the equilibrium $M_z$ in zero applied field.

The temperature dependence of $\Phi_{dc}$ is shown in Fig. S1(b), with two fits (red curves) that span the low temperature spin wave regime and the high temperature critical regime. In the high temperature regime, the best fit was obtained with Curie temperature $T_c = 175.2 \pm 0.2$ K and critical exponent $\beta = 0.33 \pm 0.03$, which is in good agreement with $\beta = 0.326$ predicted by the 3D Ising model.

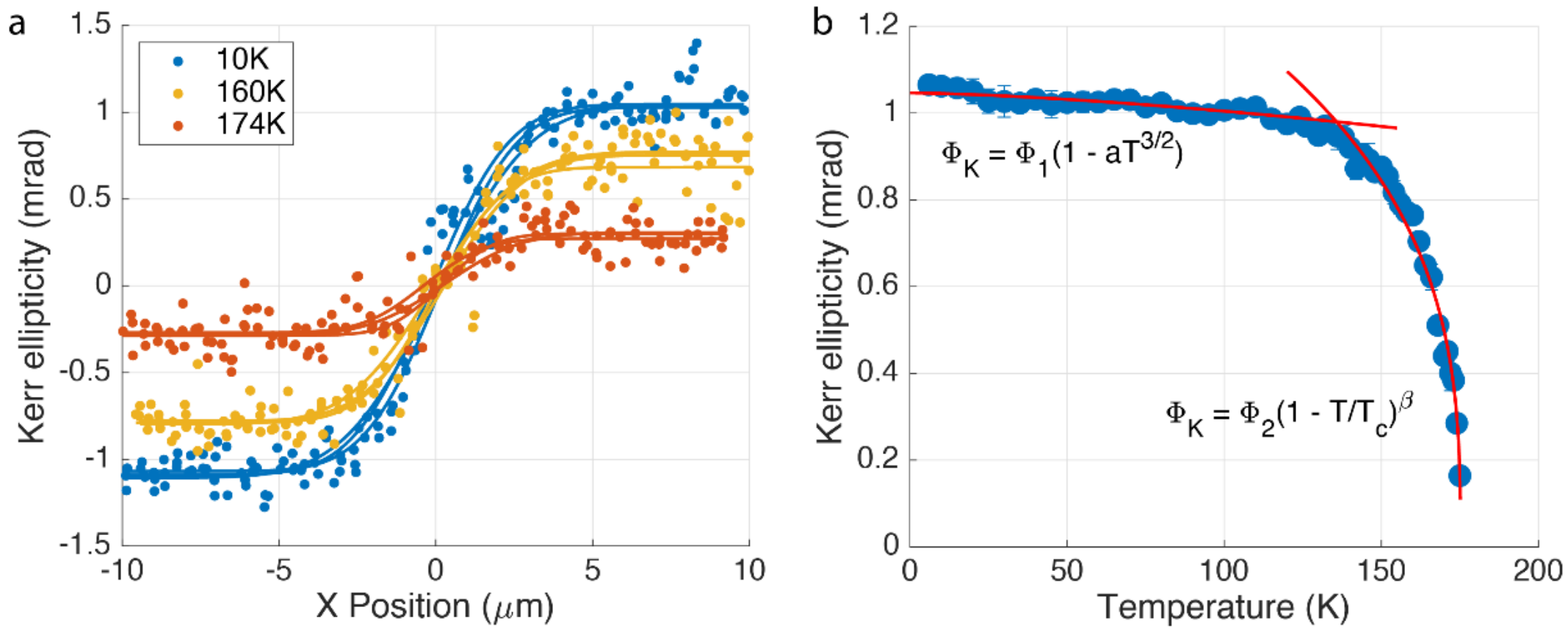

**Figure S1 | a** dc Kerr ellipticity plotted as a function of position relative to the center of the domain wall for various temperatures. The ellipticity data (dots) are subsequently fit to error functions

(lines). **b** The temperature dependence of dc Kerr ellipticity compared to predictions for the spin wave (low temperature) and critical (high temperature) regimes, respectively.

## II. Comparison of unmodulated and ac Kerr ellipticity maps

In order to verify that the ac Kerr response tracks the DW displacement, we show maps of unmodulated and ac Kerr ellipticity maps taken across the same region of the sample in Figs. S2(a) and (b), respectively.

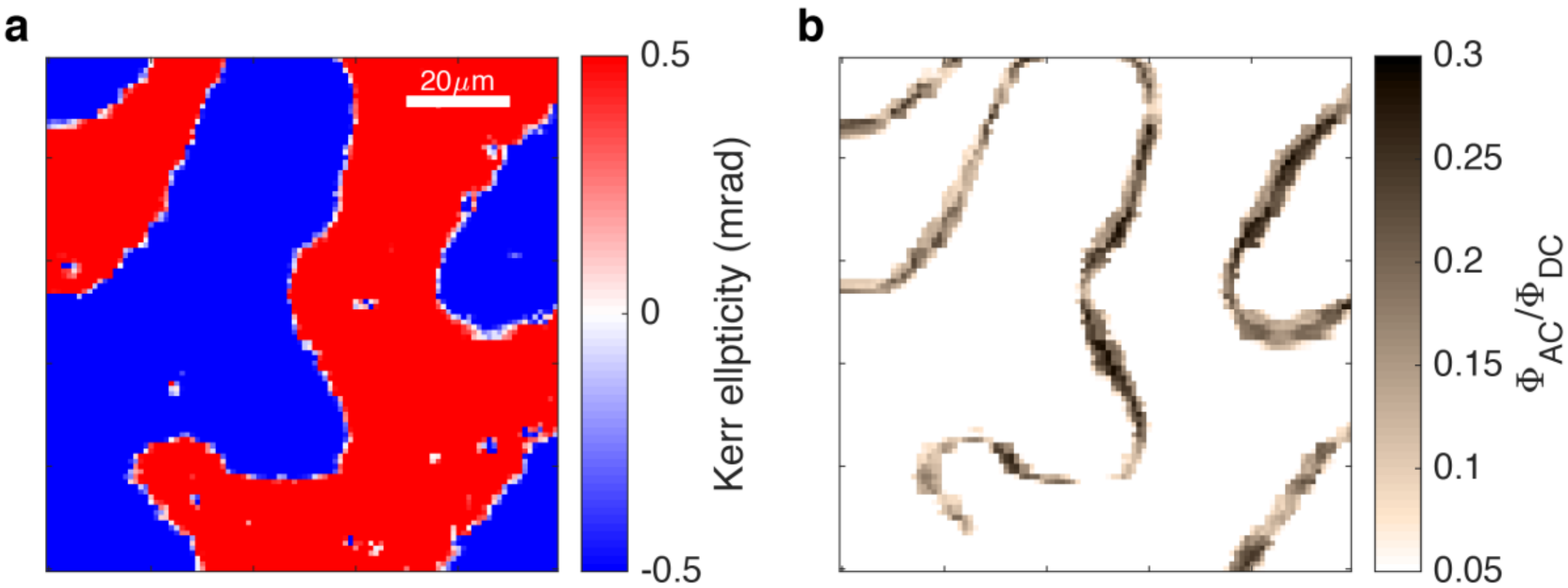

**Figure S2 | a** Unmodulated Kerr ellipticity map taken at T = 120 K reveals stripe-like magnetic domains. **b** Modulated (f = 1 kHz) ac Kerr map measured over the same region of sample exhibits a strong signal at the domain boundaries. The signal is normalized by the dc Kerr ellipticity obtained at 120 K.

## III. Analysis of domain wall displacement

Domain wall (DW) displacement was determined by measuring both $\Phi_{dc}$ and the ac modulation synchronous with the applied magnetic field, $\delta\Phi_{ac}$, as a function of position along a direction normal to the DW. Scanning the laser probe transverse to the domain wall yields a peak in $\delta\Phi_{ac}/\Phi_{dc}$ at the DW center, as discussed in the main text. There are three length scales that determine the peak amplitude: DW displacement, $\Delta x$, DW width, $w$, and laser focus radius, $\sigma$. We obtain $\sigma \approx 1\ \mu m$ using standard optical methods for determining the focal diameter. We estimate that $w$ is on the order of several nanometers based on neutron scattering experiments, where a ferromagnetic spin wave dispersion was observed, $E(q) = E_{gap} + Dq^2$, with $E_{gap} \approx 2\ meV$ and $D \approx 8\,meV - nm^2$. Thus the our measurements are performed in the regime where $\sigma \gg w$, where the peak amplitude is independent of $w$ and related to the displacement through the formula, $\delta\Phi_{ac}(0)/\Phi_{dc} = \text{erf}(\Delta x/\sqrt{2}\sigma)$.

## IV. Spatial variation of the domain wall displacement

The ac Kerr ellipticity exhibits a certain amount of spatial variation. Depending on the specific location of the measurement spot (Fig. S5a), the DW displacement can vary by a factor of two, while the threshold field $H_{th}$ exhibits a much smaller variation (Fig. S5b). However, the non-monotonic temperature dependence of the DW displacement is observed for all DWs, as shown in the ac maps plotted in Fig. 1 of the main text.

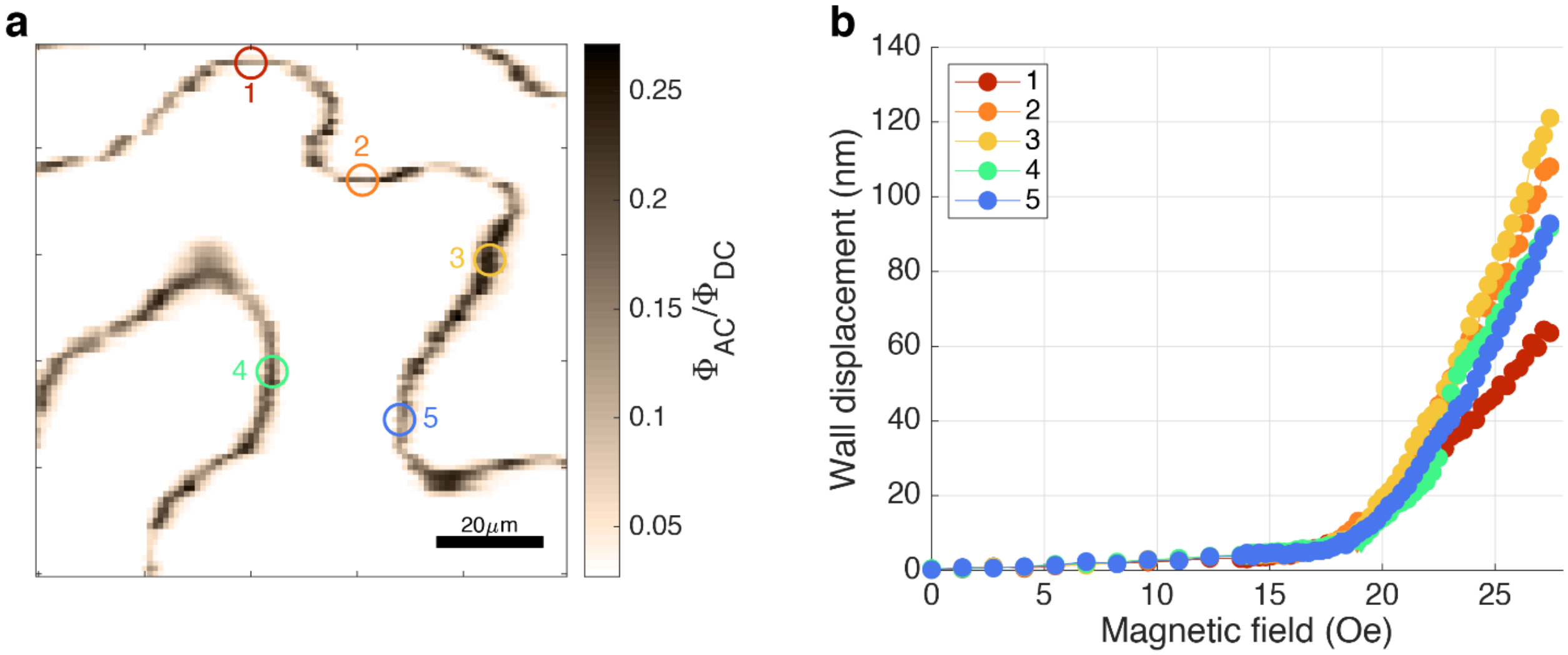

**Figure S3 | a** ac MOKE map measured at 130 K. **b** Plot of DW displacement at the five different locations indicated panel **a**, showing variations in DW mobility.

## V. Domain wall displacement at high temperatures (T > 140 K)

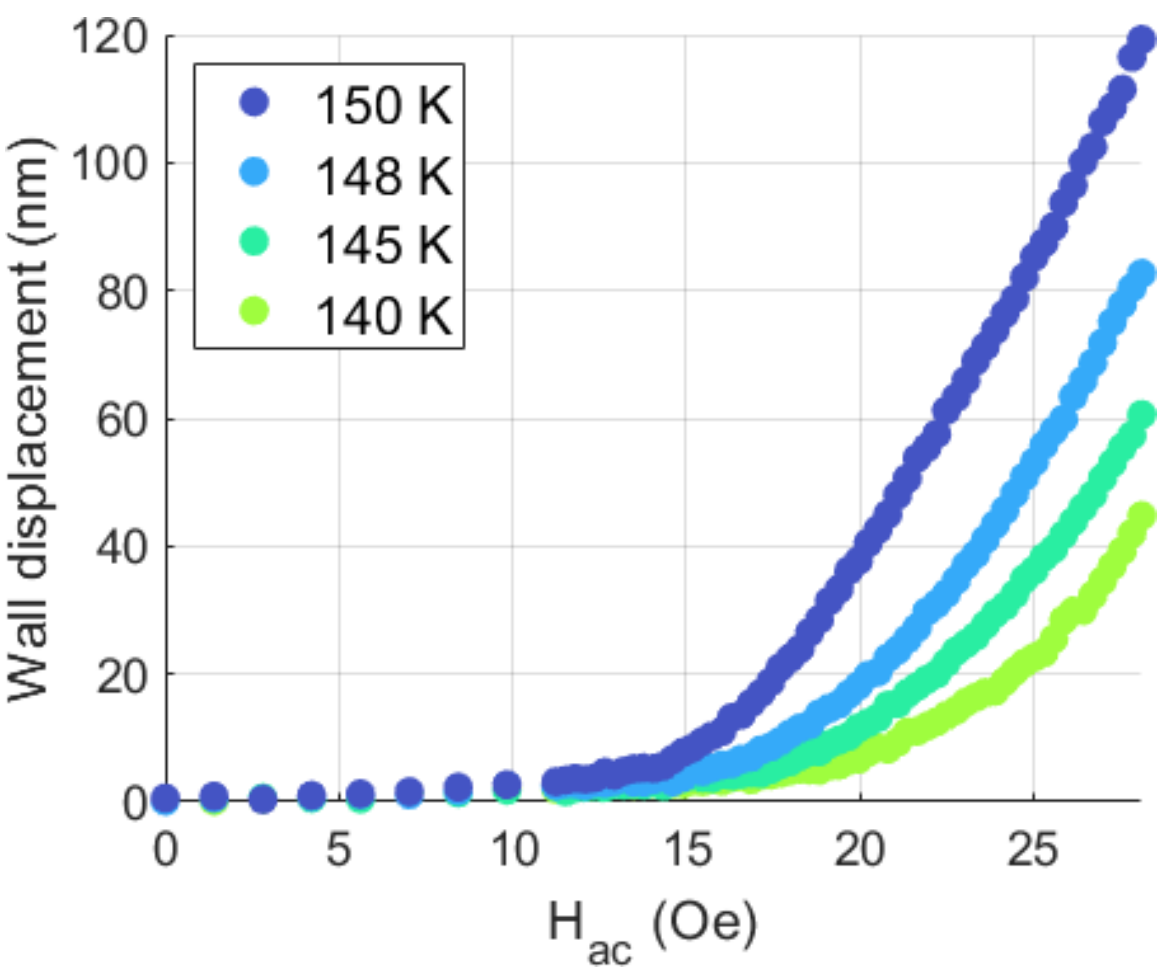

**Figure S4 | DW displacement above the BG phase transition.** Measurements of the domain wall displacement plotted against ac magnetic field amplitude at temperatures ranging from T=140 K to 150 K ($f = 1$ kHz). Larger wall displacements and mobilities can be observed toward higher temperatures.

## VI. Relation of domain wall mobility to Bulaevskii-Ginzburg order parameter

We consider the propagation of an elliptical domain wall described by magnetization,

$$\boldsymbol{M}(x,t) = M_s \left[ \hat{\boldsymbol{z}} \tanh\left(\frac{x - vt}{w}\right) + \hat{\boldsymbol{y}} \frac{\rho}{\cosh\left(\frac{x - vt}{w}\right)} \right], \tag{1}$$

where $M_s$ is the saturation magnetization, $w$ is the wall width, and $\rho = M_y/M_s$ is the normalized in-plane magnetization at the DW center. The power dissipated by the wall moving with velocity $v$ is given by,

$$P_{diss} = \frac{1}{\Gamma_L} \int dx \dot{M}^2 + \frac{1}{\Gamma_\theta} \int dx\, \dot{M}_y^2, \tag{2}$$

where $\Gamma_L$ and $\Gamma_\theta$ are relaxation rates of longitudinal and transverse fluctuations of magnetization.

Using change of variables $u = (x - vt)/w$,

$$P_{diss} = \frac{v^2}{w}\left[\frac{1}{\Gamma_L}\int du\left(\frac{dM}{du}\right)^2 + \frac{1}{\Gamma_\theta}\int du\left(\frac{dM_y}{du}\right)^2\right]. \tag{3}$$

The integral $\int du\left(\frac{dM}{du}\right)^2$ was calculated numerically. For convenience in fitting we used the empirical form $(4/3)M_s^2(1-\rho)^{2.2}$ which approximates the numerical result to within 2%. The integral $\int du\left(\frac{dM_y}{du}\right)^2$ gives $M_s^2\rho^2$, so that the dissipated power can be written,

$$P_{diss} = \frac{v^2 M_s^2}{w}\left[\frac{1}{\Gamma_L}\frac{2}{3}(1-\rho)^{2.2} + \frac{\rho^2}{\Gamma_\theta}\right]. \tag{4}$$

For steady state motion we equate the dissipated power with the rate at which stored magnetostatic energy is lowered, $U = vM_sH$, where $H$ is the magnetic field applied parallel to the easy axis. Finally solving for the ratio of the velocity to the applied field yields the relation between mobility and BG order parameter plotted in Fig. 4(c) of the main text,

$$\mu = \frac{w}{M_s}\frac{1}{\frac{2}{3\Gamma_L}(1-\rho)^{2.2} + \frac{\rho^2}{\Gamma_\theta}}.$$

## VII. Various fits to the effective DW mobility

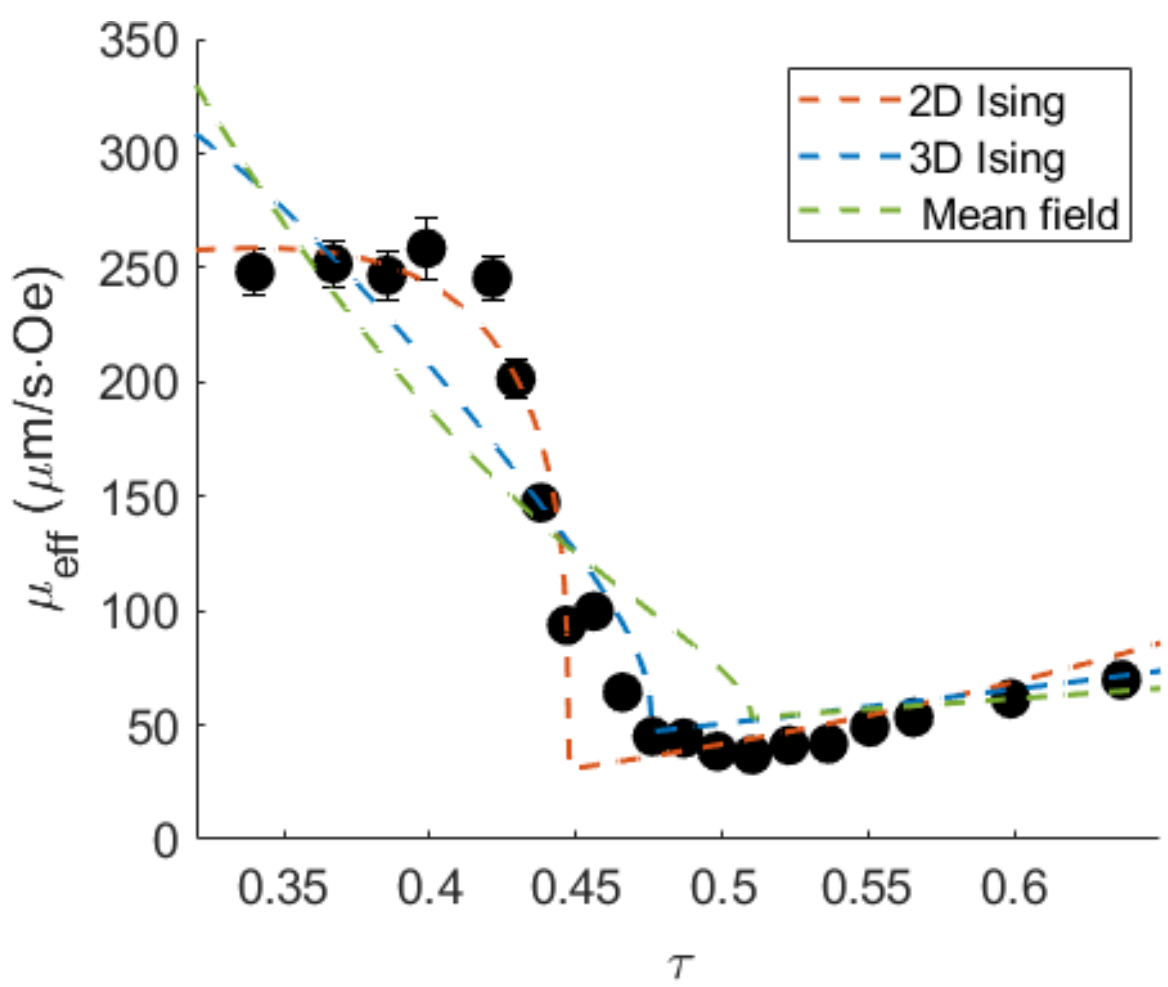

**Figure S5 | Effective DW mobility and fits to the data.** Effective mobility, $\mu_{\mathrm{eff}} = \frac{\Delta x}{\Delta H}$, of the DW is plotted against $\tau \equiv 4\chi_c/\chi_{ab}$, which is the control parameter for the linear wall phase transition. The $\mu_{\mathrm{eff}}$ data fits much better with the 2D Ising model (red dotted line, $\beta = 0.125$), compared to the 3D Ising model (blue dotted line, $\beta = 0.326$) or the mean field theory (green dotted line, $\beta = 0.5$) with critical values of $\tau$ near 0.5. In all cases, $\tau^*$ was adjusted give the best fit.

### References

1 Garanin, D. A. Dynamics of elliptic domain walls. *Physica A: Statistical Mechanics and its Applications* **178**, 467-492, doi:10.1016/0378-4371(91)90033-9 (1991).